\documentclass[10pt,conference]{IEEEtran}

\usepackage{graphicx}
\usepackage{fancyhdr}					
\usepackage{amsmath,amssymb,amsfonts}   
\usepackage{glossaries}                 
\usepackage{enumitem}                   
\usepackage{makecell}                   
\usepackage{pifont}                     
\usepackage{cleveref}                   
\usepackage{color-edits}                
\usepackage{multirow}                   
\usepackage[table]{xcolor}              
\usepackage{float}                      
\usepackage{caption}                    
\usepackage{booktabs}                   
\usepackage{balance}                    
\usepackage{setspace}                   
\usepackage{xspace}                     
\usepackage{algorithm}                  
\usepackage{algpseudocode}              
\usepackage{tcolorbox}					

\definecolor{lightgray}{HTML}{CCCCCC}

\newif\ifnb     
\nbfalse
\nbtrue

\newif\ifcom    
\comtrue

\newif\ifarx    
\arxfalse

\newcommand{\name}{LASH-PAT\xspace}

\crefname{algocf}{Alg.}{Algs.}
\crefname{algorithm}{Alg.}{Algs.}
\crefname{figure}{Fig.}{Figs.}
\crefname{equation}{Eq.}{Eqs.}

\addauthor[Patrick]{pi}{teal}
\renewcommand{\pi}[1]{\ifcom\picomment{#1}\fi}
\definecolor{darkred}{rgb}{0.5,0,0}
\definecolor{darkgreen}{rgb}{0,0.5,0}
\newcommand{\rd}[1]{\ifcom\textcolor{darkred}{#1}\fi}
\newcommand{\gn}[1]{\ifcom\textcolor{darkgreen}{#1}\fi}

\newcommand{\cmark}{\ding{52}}%
\newcommand{\xmark}{\ding{56}}%
\newcommand{\rone}{\raisebox{-0.125em}{\includegraphics[height=0.75em]{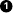}}}
\newcommand{\rtwo}{\raisebox{-0.125em}{\includegraphics[height=0.75em]{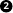}}}
\newcommand{\rthree}{\raisebox{-0.125em}{\includegraphics[height=0.75em]{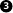}}}
\newcommand{\rfour}{\raisebox{-0.125em}{\includegraphics[height=0.75em]{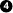}}}
\newcommand{\rfive}{\raisebox{-0.125em}{\includegraphics[height=0.75em]{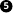}}}

\newcommand{\reight}{\raisebox{-0.125em}{\includegraphics[height=0.75em]{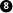}}}

\newcommand{\rsixteen}{\raisebox{-0.125em}{\includegraphics[height=0.75em]{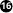}}}
\newcommand{\rseventeen}{\raisebox{-0.125em}{\includegraphics[height=0.75em]{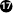}}}
\newcommand{\reighteen}{\raisebox{-0.125em}{\includegraphics[height=0.75em]{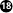}}}
\newcommand{\rnineteen}{\raisebox{-0.125em}{\includegraphics[height=0.75em]{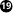}}}
\newcommand{\rtwenty}{\raisebox{-0.125em}{\includegraphics[height=0.75em]{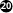}}}

\newcommand{\rtwentyone}{\raisebox{-0.125em}{\includegraphics[height=0.75em]{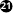}}}
\newcommand{\rtwentytwo}{\raisebox{-0.125em}{\includegraphics[height=0.75em]{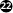}}}
\newcommand{\rtwentythree}{\raisebox{-0.125em}{\includegraphics[height=0.75em]{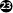}}}
\newcommand{\rtwentyfour}{\raisebox{-0.125em}{\includegraphics[height=0.75em]{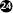}}}
\newcommand{\rtwentyfive}{\raisebox{-0.125em}{\includegraphics[height=0.75em]{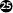}}}
\newcommand{\rtwentysix}{\raisebox{-0.125em}{\includegraphics[height=0.75em]{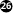}}}
\newcommand{\rtwentyseven}{\raisebox{-0.125em}{\includegraphics[height=0.75em]{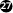}}}

\newcommand{\rfifty}{\raisebox{-0.125em}{\includegraphics[height=0.75em]{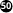}}}

\newtcbox{\gray}[1][]{%
  on line,
  colback=gray!30,   
  colframe=gray!30,  
  boxrule=0pt,       
  arc=3pt,           
  boxsep=0pt,        
  left=1pt,
  right=1pt,
  top=1pt,
  bottom=1pt,
  #1
}

\newcommand{\githuburl}{\ifnb https://github.com/spcl/LASH-PAT\else https://github.com/anon8274/lash\_pat (anonymized for review)\fi}

\newacronym{d2d}{D2D}{die-to-die}
\newacronym{ci}{CI}{confidence interval}
\newacronym{llm}{LLM}{large language model}
\newacronym{moe}{MoE}{Mixture of Experts}
\newacronym{fft}{FFT}{fast Fourier transform}
\newacronym{phy}{PHY}{physical layer}
\newacronym{noc}{NoC}{network-on-chip}
\newacronym{ici}{ICI}{inter-chiplet interconnect}
\newacronym{pici}{ICI-P}{inter-chip interconnect with a passive substrate}
\newacronym{aici}{ICI-A}{inter-chip interconnect with an active interposer}
\newacronym{rf}{RF}{routing function}
\newacronym{sf}{SF}{selection function}
\newacronym{vc}{VC}{virtual channel}
\newacronym{ml}{ML}{machine learning}
\newacronym{flit}{flit}{flow control unit}
\newacronym{cdg}{CDG}{channel dependency graph}
\newacronym{mst}{MST}{minimum spanning tree}

\newacronym{scb}{SCB}{simple cycle breaking}
\newacronym{dl}{DL}{descending layer}

\title{LASH-PAT: Deadlock-Free Adaptive Shortest-Path Routing for Inter-Chiplet Interconnects}

\ifnb
	
\fi

\author{
  \ifnb
    \IEEEauthorblockN{Patrick Iff and Torsten Hoefler}
    \IEEEauthorblockA{ETH Zurich, Zurich, Switzerland \\ \{patrick.iff, torsten.hoefler\}@inf.ethz.ch}
  \else
    \IEEEauthorblockN{Anonymous Author(s)}
	\IEEEauthorblockA{{~}\\{~}}
  \fi
}

\begin{document}

\maketitle

\begin{abstract}
	With the slowdown of transistor scaling, chiplet-based systems, and in particular chiplet-based wafer-scale integration, have emerged as a promising approach to sustain growth in computing capability per cost for modern processors and accelerators.
In recent years, many specialized inter-chiplet interconnect topologies have been proposed.
In these topologies, the simple, well-established, and deadlock-free dimension-order routing scheme is not applicable due to its reliance on regular mesh structures.
In this work, we study routing in these recently proposed inter-chiplet interconnect topologies by categorizing and evaluating nine deadlock-free routing functions for irregular topologies known from data center networking in this new context.
This reveals a trade-off between turn-based routing, which enables path diversity, and layered routing, which ensures shortest-path delivery.
To resolve this trade-off, we propose \name, a routing function that extends layered shortest-path routing with adaptive path diversity.
\name supports multiple shortest paths and dynamically selects among them based on local congestion while preserving deadlock freedom.
When compared to the best-performing baseline routing function for each combination of topology, traffic pattern, and evaluated metric, LASH-PAT improves throughput by 21.15\% and reduces energy per packet by 12.77\%, while providing latency comparable to the lowest-latency baseline routing function.

\end{abstract}

\section{Introduction}
\label{sec:intro}

The growing compute demands of increasingly large \gls{ml} models require continuous improvements in the computing capability per cost of high-performance processors and accelerators.
For multiple decades, the computer architecture community was able to meet this demand by scaling up the number of transistors per chip, enabled by Moore's law and Dennard scaling.
However, due to the slowdown of Moore's law and the end of Dennard scaling, new approaches are required to sustain performance improvements.
The two most promising directions are 2.5D chiplet integration \cite{chiplet-1,chiplet-2,occamy} and wafer-scale integration \cite{wsi-1,wsi-2,wsi-3,wsi-4,nw-for-wsi}.
The former reduces design and manufacturing cost by enabling system modularity, chiplet reuse, heterogeneous integration, frequency and power binning, and improved yield \cite{chiplet-survey,amd-chiplets}, while the latter enables packing more compute power into the high-bandwidth on-chip communication domain \cite{wsi-survey}.
When combined, these two approaches are referred to as chiplet-based wafer-scale integration \cite{chiplet-based-wsi}.

Since data movement is a major bottleneck for many \gls{ml} workloads \cite{data-movement}, the design of scalable and efficient \gls{ici} is critical to overall system performance.
While the 2D mesh topology has long been the de facto standard for \glspl{noc}, recent work has proposed specialized \gls{ici} topologies \cite{hexamesh,foldedhexatorus,double-butterfly,butterdonut,kite,sid-mesh,cluscross} for both 2.5D chiplet systems on passive substrates \cite{glass-1,silicon-vs-organic,interposer-tech} and active interposers \cite{intact,active-interposer}.
These topologies exploit the unique characteristics of 2.5D chiplet systems to improve performance, but they no longer support the established deadlock-free dimension-order routing algorithm, which relies on a mesh structure.

To effectively utilize these novel topologies, it is necessary to understand and evaluate deadlock-free routing algorithms for irregular networks.
In this work, we investigate the performance of nine different \glspl{rf} for irregular topologies \cite{lash,updown,prefix-routing,lturn,scb,dfsssp,lash-tor,a-lash}, primarily known from data center networking, when applied to chiplet-based wafer-scale systems.
These methods can be broadly divided into two categories based on their deadlock-avoidance mechanism.
\emph{Turn-based routing} is similar to turn models for 2D mesh topologies \cite{tm,oetm} and achieves deadlock freedom by prohibiting specific ``turns'' in the network.
\emph{Layered routing} achieves deadlock freedom by using virtual layers to break cyclic channel dependencies.
Turn-based \glspl{rf} typically provide path diversity but do not necessarily provide shortest paths, while layered \glspl{rf} often guarantee shortest-path routing but may not provide path diversity.
Exploiting path diversity is important to mitigate local congestion and achieve high throughput, while shortest-path routing is important to minimize latency.

To combine the benefits of turn-based and layered routing, we propose \name, a novel \gls{rf} that extends \underline{la}yered \underline{sh}ortest-path routing with \underline{pat}h diversity.
\name builds upon the LASH-TOR \gls{rf} \cite{lash-tor} through two key mechanisms:
\gray{\textbf{M1}}: augmenting routing tables with multiple next-hop candidates to enable adaptive routing for congestion mitigation, and
\gray{\textbf{M2}}: enabling fine-grained path-to-layer assignments to minimize the number of layers required for deadlock-free shortest-path routing.
Evaluated against nine \glspl{rf} for irregular topologies, \name achieves 31.92\% higher throughput and 16.61\% lower energy per packet than the highest-throughput baseline (SCB \cite{scb}), while exhibiting latency comparable to the lowest-latency baseline (DFSSSP \cite{dfsssp}).

In summary, the main contributions of this paper are:
\begin{itemize}
\item We present the first comprehensive performance investigation of \glspl{rf} for irregular topologies in the context of inter-chiplet interconnects.
\item We propose and evaluate \name, an \gls{rf} for irregular topologies that extends LASH routing with path diversity.
\end{itemize}

\newpage
\section{Background}
\label{sec:back}

\glsresetall

\subsection{2.5D Chiplet-Based Systems}
\label{sec:back-chiplet}

A 2.5D chiplet-based system consists of multiple silicon chiplets that are attached to and connected through a common substrate.
This substrate can be passive, containing only wires without repeaters, or active, integrating repeaters and potentially routers and power delivery components.
Common passive substrates include established organic substrates \cite{organic-interposer}, passive silicon interposers \cite{si-interposer}, silicon bridges \cite{emib,dbhi}, and emerging glass substrates \cite{glass-1}, while active interposers \cite{active-interposer,intact} are fabricated in silicon using processes similar to those of the chiplets themselves.
Chiplets are connected through \gls{d2d} channels \cite{link-1,link-2}, which require a \gls{phy} \cite{phy-1} at both the sending and the receiving chiplet to translate between on-chip and inter-chiplet interconnect protocols \cite{bow,ucie,axi,chi}.
Since this work focuses on the design of the \gls{ici}, we abstract the internal \gls{noc} of each chiplet as a single router, following the approach of Yin et al. \cite{abstract-router-model}, who model the \gls{ici} as a single router and focus on the design of the \gls{noc}.

\subsection{Inter-Chiplet Interconnection Networks}
\label{sec:back-ici}

We assume a packet-switched network with input-queued routers, wormhole routing, and credit-based flow control, as commonly used in on-chip and inter-chiplet networks.
Each message is divided into packets, and each packet is further divided into \glspl{flit}.
Packets of the same message may take different paths through the network (packet switching), whereas flits within a packet follow a single path in order and are not interleaved with flits from other packets (wormhole routing).
Routers advertise available input buffer space to their neighbors using credits, where one credit corresponds to one flit of buffer capacity and a flit can be forwarded from router A to router B if router A receives sufficient credits from router B, indicating that router B has enough buffer space to accept the flit (credit-based flow control).
This lossless flow control scheme avoids costly end-to-end acknowledgments and retransmissions.
A key but often underutilized benefit of credit-based flow control is that each router has local visibility of downstream buffer occupancy and can therefore make informed routing decisions based on congestion information.
This property is leveraged by our key mechanism \gray{\textbf{M1}} to enable adaptive routing.

Each physical channel is logically partitioned into multiple \glspl{vc}, each with a dedicated input buffer.
Virtual channels decouple buffer resources from channel resources, allowing packets that are waiting for a busy channel to be overtaken by packets that are ready to advance, thereby improving throughput.
Virtual channels can also be used to implement deadlock-avoidance schemes.

\subsection{Deadlock-Free Routing}
\label{sec:back-routing}

In this work, we assume distributed routing, where each time a packet arrives at a router, the router invokes a routing algorithm to determine the output port and \gls{vc} for the packet's next hop.
The routing algorithm consists of a \gls{rf} that determines a set of candidate (out\_port,out\_vc)-pairs, and a \gls{sf} that selects one of these candidates.
The \gls{rf} must ensure deadlock freedom, while the \gls{sf} is responsible for avoiding congestion.

To reason about the deadlock freedom of an \gls{rf}, it is useful to consider the \gls{cdg}.
The \gls{cdg} contains one vertex per channel and an edge between channels A and B if the \gls{rf} allows a packet arriving on channel A to be forwarded to channel B.
If the \gls{cdg} is acyclic, the \gls{rf} is guaranteed to be deadlock-free \cite{cdg-1,cdg-2,cdg-3}.
To accommodate \glspl{vc}, one can either construct one \gls{cdg} per \gls{vc} and ensure that packets only transition from lower- to higher-numbered \glspl{vc}, or construct a single \gls{cdg} with one vertex per virtual channel, as demonstrated by Pascual et al. \cite{vc-cdg}.
Our key mechanism \gray{\textbf{M2}} is based on the latter approach, enabling more flexible path-to-layer assignment than existing \glspl{rf}.

\section{Related Work}
\label{sec:rl}

\begin{table*}[h]
\footnotesize
\setlength{\tabcolsep}{5.5pt}
\centering
\captionsetup{justification=centering}
\caption{(\textsection \ref{sec:rl-irregular}) \textbf{Overview of routing functions for irregular network topologies}.
$^\dagger$Uses a turn-based routing function as a fallback if the number of available virtual channels is insufficient, which may result in non-shortest paths.
$^\ddagger$In addition to the routing table, this algorithm uses a layer table mapping destinations to virtual layers or another auxiliary data structure.
$^*$Only allows transitions to higher/lower layers.
$^{**}$Provides adaptivity only at the source router.
}
\vspace{-0.75em}
\begin{tabular}{lcccccccll}
\toprule
\textbf{\makecell[l]{Name}}
	& \rotatebox[origin=l]{90}{\textbf{\makecell[l]{Unrestricted\\use of\\physical channels}}}
	& \rotatebox[origin=l]{90}{\textbf{\makecell[l]{Unrestricted\\use of\\virtual channels}}}
	& \rotatebox[origin=l]{90}{\textbf{\makecell[l]{Supports path\\diversity w.r.t.\\physical channels}}}
	& \rotatebox[origin=l]{90}{\textbf{\makecell[l]{Supports path\\diversity w.r.t.\\virtual channels}}}
	& \rotatebox[origin=l]{90}{\textbf{\makecell[l]{Routing\\over shortest\\paths only}}}
	& \rotatebox[origin=l]{90}{\textbf{\makecell[l]{Allows transition\\between\\virtual layers}}}
	& \rotatebox[origin=l]{90}{\textbf{\makecell[l]{Supports\\networks without\\virtual channels}}}
	& \rotatebox[origin=l]{0}{\textbf{\makecell[l]{Routing table index}}}
	& \rotatebox[origin=l]{0}{\textbf{\makecell[l]{Routing table entry}}}
\\
\midrule
	Prefix \cite{prefix-routing} 	& \xmark	& \cmark	& \xmark		& \cmark		& \xmark			& \cmark	& \cmark			& destination					& out\_port					\\
	Up*/Down* \cite{updown} 			& \xmark	& \cmark	& \cmark		& \cmark		& \xmark			& \cmark	& \cmark			& destination, in\_port			& list(out\_port)			\\
	Left/Right \cite{lturn} 		& \xmark	& \cmark	& \cmark		& \cmark		& \xmark			& \cmark	& \cmark			& destination, in\_port			& list(out\_port)			\\
	L-Turn \cite{lturn} 			& \xmark	& \cmark	& \cmark		& \cmark		& \xmark			& \cmark	& \cmark			& destination, in\_port			& list(out\_port)			\\
	SCB \cite{scb}  				& \xmark	& \cmark	& \cmark		& \cmark		& \xmark			& \cmark	& \cmark			& destination, in\_port			& list(out\_port)			\\
	LASH \cite{lash} 				& \xmark	& \xmark	& \xmark		& \xmark		& \cmark			& \xmark	& \xmark			& destination$^\ddagger$		& out\_port$^\ddagger$		\\
	A-LASH \cite{a-lash}			& \xmark	& \xmark	& \cmark		& \cmark$^{**}$	& \cmark			& \xmark	& \xmark			& destination$^\ddagger$		& list(out\_port)$^\ddagger$\\
	MP-LASH \cite{a-lash}			& \xmark	& \xmark	& \cmark$^{**}$	& \cmark$^{**}$	& \cmark$^\dagger$	& \xmark	& \cmark$^\dagger$	& address (mapped to a dst)		& out\_port					\\	
	DFSSSP \cite{dfsssp} 			& \xmark	& \xmark	& \xmark		& \xmark		& \cmark			& \xmark	& \xmark			& destination$^\ddagger$		& out\_port$^\ddagger$		\\		
	Nue \cite{nue}					& \xmark	& \xmark	& \xmark		& \xmark		& \cmark$^\dagger$	& \xmark	& \cmark$^\dagger$	& destination, in\_port			& out\_port					\\
	LASH-TOR \cite{lash-tor} 		& \xmark	& \xmark	& \xmark		& \xmark		& \cmark$^\dagger$	& \cmark$^*$& \cmark$^\dagger$	& destination, in\_port, in\_vc	& out\_port, out\_vc		\\
	DL \cite{dl}					& \xmark	& \xmark	& \xmark		& \xmark		& \xmark			& \cmark$^*$& \cmark$^\dagger$	& destination, in\_port, in\_vc	& out\_port, out\_vc		\\
	Sancho et al. \cite{sancho}		& \xmark	& \xmark	& \xmark		& \xmark		& \cmark$^\dagger$ 	& \cmark$^*$& \cmark$^\dagger$	& destination$^\ddagger$		& out\_port$^\ddagger$		\\
\midrule
	\textbf{LASH-PAT}				& \xmark	& \xmark	& \cmark		& \cmark		& \cmark$^\dagger$	& \cmark	& \cmark$^\dagger$	& destination, in\_port, in\_vc	& list(out\_port, out\_vc)	\\
\bottomrule
\end{tabular}
\label{tab:rf}
\vspace{-2.25em}
\end{table*}

\subsection{Routing in 2D Mesh Networks}
\label{sec:rl-mesh}

Research on on-chip routing has primarily focused on the established 2D mesh topology, which is widely used in commercial products \cite{dojo} and has been extensively studied in the literature \cite{mesh-survey-1,mesh-survey-2}.
\glspl{rf} for 2D mesh \glspl{noc} can achieve deadlock freedom through different forms of channel restriction.  
One approach restricts the use of physical channels while allowing all \glspl{vc} to be used, as seen in turn-based \glspl{rf} for irregular topologies.  
Another approach restricts a combination of physical and virtual channels, as seen in most layered \glspl{rf} for irregular topologies.  
A third approach restricts only virtual channels while allowing all physical channels to be used, which is rare in \glspl{rf} for irregular topologies.
The established dimension-order routing (DOR), the Turn Model \cite{tm}, the Odd-Even Turn Model \cite{oetm}, the Abacus Turn Model \cite{abtm}, and the Direction Restriction Model \cite{drm} all only restrict the use of physical channels.
MVI \cite{mvi}, 3P \cite{3p}, ESPADA \cite{espada}, and MESH\_ROUTE \cite{mr} only restrict the use of virtual channels.
P-Adaptive \cite{padaptive} and LEF \cite{lef} restrict the use of both physical and virtual channels and are therefore conceptually most similar to layered \glspl{rf} for irregular topologies, including our novel \name \gls{rf}.

\glspl{sf} for 2D mesh \glspl{noc} can be oblivious and not consider congestion information, such as Z$^2$ \cite{z2}, or they can use congestion information from various scopes.
PCA \cite{pca}, DyXY \cite{dyxy}, PCAR \cite{pcar}, and the \glspl{sf} proposed by Shafiei et al. \cite{shafiei} and Nickray et al. \cite{nickray} use congestion information from neighboring routers, which, in networks with credit-based flow control, is available in the form of free buffer slots.
NoP \cite{nop}, RCA \cite{rca}, DBAR \cite{dbar}, and CATRA \cite{catra} gather congestion information from larger network regions, while DAR \cite{dar} and DPN \cite{dpn} leverage global congestion information.

Other works on on-chip routing include Multi \cite{multi}, which guarantees in-order delivery while using multiple paths, DyAD \cite{dyad}, which switches between deterministic and adaptive routing based on congestion, as well as many works on fault-tolerant routing \cite{ftr-1,ftr-2,ftr-3,ftr-4,ftr-5,ftr-6} and application-specific routing \cite{asr-1,asr-2}.

\subsection{Routing in Irregular Networks}
\label{sec:rl-irregular}
\vspace{-0.1em}

Routing functions for irregular networks can be categorized into two groups: turn-based \glspl{rf} and layered \glspl{rf}.

\underline{Turn-based \glspl{rf}}, such as Prefix routing \cite{prefix-routing}, Up*/Down* \cite{updown}, Left/Right \cite{lturn}, L-Turn \cite{lturn}, and the \gls{scb} algorithm \cite{scb}, restrict specific ``turns,'' i.e., combinations of incoming and outgoing channels, to break cycles in the \gls{cdg}.
They neither require nor restrict \glspl{vc}.
Except for Prefix routing, all turn-based \glspl{rf} support multiple next-hop candidates, which is a key advantage over layered \glspl{rf}, as it provides path diversity and enables adaptive routing, thereby improving throughput.
The drawback of turn-based \glspl{rf} is that the turn restrictions often lead to non-shortest paths in many topologies, as we show in \Cref{sec:eval}.

\underline{Layered \glspl{rf}}, such as LASH \cite{lash}, LASH-TOR \cite{lash-tor}, DFSSSP \cite{dfsssp}, and Nue \cite{nue}, search for shortest paths in the network and avoid deadlocks by partitioning these paths into virtual layers (implemented using \glspl{vc}) such that the \gls{cdg} of each layer is acyclic.
In DFSSSP, LASH, and Nue, the source router contains a layer table that maps each destination address to a layer, and once in flight, a packet is not allowed to change its layer.
LASH-TOR (TOR denotes transition-oriented routing) allows packets to move to higher layers, enabling more flexible path assignment while ensuring deadlock freedom.

While our proposed \name \gls{rf} builds on the principles of LASH-TOR, it introduces two key mechanisms to improve performance:
\gray{\textbf{M1}}: Routing tables store multiple (out\_port, out\_vc) pairs per entry.
This allows the \gls{sf} to dynamically select from multiple next-hop candidates, enabling adaptive routing and improving throughput.
\gray{\textbf{M2}}: We define the \gls{cdg} with one vertex per \gls{vc}, allowing packets to transition across layers, not only to higher ones, provided that the \gls{cdg} remains acyclic.
This flexible layering scheme reduces the number of \glspl{vc} required to guarantee deadlock freedom.
In conclusion, \gray{\textbf{M1}} combines the path diversity of turn-based routing with the shortest-path properties of layered routing, while \gray{\textbf{M2}} improves \gls{vc} efficiency and reduces hardware overhead.
\Cref{tab:rf} compares \name with existing irregular-topology \glspl{rf}.

We are not the first to extend layered \glspl{rf} for irregular topologies with path diversity.
MP-LASH and A-LASH \cite{a-lash} are two such \glspl{rf} proposed for InfiniBand networks \cite{ib1,ib2}, but both rely on an InfiniBand feature that allows multiple addresses per destination, which is not supported in standard \gls{ici} architectures.
We implement A-LASH using only standard \gls{ici} features, whereas this is not possible for MP-LASH.
MP-LASH assigns one address per unique shortest path between a given source and destination.
By selecting among multiple addresses that map to the same destination, the source router can choose a specific path per packet.
Once in flight, a packet's path is fixed, so there is no adaptivity beyond the source router.
A-LASH instead assigns one address per source-attached channel and allows packets sent to this address to use any shortest path starting with the corresponding channel.
Unlike MP-LASH, this enables adaptive routing throughout the network, but requires assigning all shortest paths starting from the same channel to a single layer without introducing cycles in that layer's \gls{cdg}, which can require a large number of \glspl{vc}.
LASH-PAT differs significantly from both schemes, as it enables adaptive routing throughout the network (unlike MP-LASH, via \gray{\textbf{M1}}) while enabling fine-grained path assignment to layers (unlike A-LASH, via \gray{\textbf{M2}}).

Two additional \glspl{rf} that combine concepts of layered and turn-based routing are \gls{dl} routing \cite{dl}, which uses a separate turn-based \gls{rf} per layer while allowing layer transitions only to lower layers, and the work by Sancho et al. \cite{sancho}, which is based on Up*/Down* routing but allows packets to take otherwise forbidden turns as long as they transition to a higher layer when taking such a turn.

\section{The LASH-PAT Routing Function}
\label{sec:lash-pat}

\subsection{Description of LASH-PAT}
\label{sec:lash-pat-desc}

\subsubsection{Motivation}
\label{sec:lash-pat-motivation}

The goal of LASH-PAT is to extend layered routing, which provides shortest paths and uses virtual layers for deadlock avoidance, via two key mechanisms:
\gray{\textbf{M1}}: providing multiple shortest paths per destination, enabling congestion-aware selection and improving throughput.
\gray{\textbf{M2}}: using the \gls{cdg} over \glspl{vc} as defined by Pascual et al. \cite{vc-cdg} for deadlock avoidance, which reduces the number of layers and hence the number of \glspl{vc} required compared to prior work.

\subsubsection{Deployment}
\label{sec:lash-pat-deploy}

LASH-PAT routing consists of an offline phase that computes routing tables and an online phase that uses them to forward packets through the network.

\underline{In the offline phase}, the routing tables for each router in the network are computed at design time.
In chiplet-based systems that are guaranteed to be free of faults, which can, e.g., be achieved by known-good-die testing, the routing tables can be computed once and written into the routers of each chip.
If static faults are possible, the chiplet-based system must be tested to identify the fault configuration at hand, and the LASH-PAT algorithm is executed once per distinct fault configuration to compute the routing tables for each router in the network.
Since LASH-PAT is applicable to arbitrary topologies, it can handle any fault configuration as long as the network remains connected.
The LASH-PAT routing tables are indexed by the destination router ID, the incoming port ID, and the incoming \gls{vc} ID, and they contain a list of outgoing port ID and outgoing \gls{vc} ID pairs.

\underline{In the online phase}, whenever a header flit arrives at a router, the router looks up the routing table and forwards the list of possible next hops to the selection function, which then selects one of the next hops based on network congestion.

\subsubsection{LASH-PAT Algorithm to Build the Routing Tables}
\label{sec:lash-pat-algorithm}

Building the routing tables for LASH-PAT is divided into two steps, as outlined\footnote{All algorithm descriptions are provided in pseudocode to keep the description concise and to focus on the algorithmic ideas. A fully working C++ implementation of the LASH-PAT algorithm is available in our open-source repository: \githuburl.} in \Cref{alg:build-routing-tables}, which calls \Cref{alg:configure-custom-base-routing,alg:configure-fallback-base-routing,alg:improve-fallback-base-routing,alg:insert-additional-paths,alg:find-path}.

\underline{In the first step}, we configure a deterministic base routing by finding exactly one deterministic path from each source to each destination (see \Cref{alg:configure-custom-base-routing}).
This is done using a modified version of Dijkstra's algorithm (see \Cref{alg:find-path}), which only allows paths that do not create cycles in the \gls{cdg} over \glspl{vc} (\gray{\textbf{M2}}), and, instead of using unit weights for edges, favors edges that do not induce additional edges in the \gls{cdg}.
This increases the chance of successfully configuring the base routing as well as the chance of finding additional paths in the second step.
Note that while this favors shortest paths, it may result in non-shortest paths if it is otherwise not possible to find a path that does not create cycles in the \gls{cdg}.
This step can fail if there are not enough \glspl{vc} available to guarantee deadlock freedom in a given network topology.
If this step fails, we use a turn-based fallback routing function to configure a deterministic base routing, which may also lead to non-shortest paths (see \Cref{alg:configure-fallback-base-routing}).
We then use our modified version of Dijkstra's algorithm to try to improve the base routing provided by the fallback routing function by replacing some of its paths with shorter ones (see \Cref{alg:improve-fallback-base-routing}).

\algnewcommand{\Global}[1]{\Statex \hspace{-1.7em}\textbf{Global} #1}

\begin{algorithm}[h]
\small
\caption{\textbf{(\textsection \ref{sec:lash-pat-algorithm}) BuildRoutingTables}}
\label{alg:build-routing-tables}
\begin{algorithmic}[1]
\Require n\_additional\_paths, n\_routers
\Global routing\_tables, CDG

	\hspace{-3em}\Comment{\parbox[t]{0.97\linewidth}{Step 1: Configuring deterministic base routing}}
	\State success = ConfigureCustomBaseRouting(n\_routers) 	\Comment{Alg. \ref{alg:configure-custom-base-routing}}
	\If{not success}
		\State Reset routing\_tables and CDG
		\State ConfigureFallbackBaseRouting(n\_routers)			\Comment{Alg. \ref{alg:configure-fallback-base-routing}}
		\State ImproveFallbackBaseRouting(n\_routers)			\Comment{Alg. \ref{alg:improve-fallback-base-routing}}
	\EndIf

	\hspace{-3em}\Comment{\parbox[t]{0.97\linewidth}{Step 2: Configuring additional paths}}
	\For{i = 1 to n\_additional\_paths}
		\State success = InsertAdditionalPaths(n\_routers)		\Comment{Alg. \ref{alg:insert-additional-paths}}
		\If{not success}
			\State break
		\EndIf
	\EndFor
\end{algorithmic}
\end{algorithm}

\vspace{-1.25em}
\begin{algorithm}[h]
\small
\caption{\textbf{(\textsection \ref{sec:lash-pat-algorithm}) ConfigureCustomBaseRouting}}
\label{alg:configure-custom-base-routing}
\begin{algorithmic}[1]
\Require{n\_routers}
	\For{src = 1 to n\_routers in random order}
		\For{dst = 1 to n\_routers in random order}
			\State success = FindPath(src, dst, $\infty$) \Comment{Alg. \ref{alg:find-path}}
			\If{not success}
				\State \textbf{return} false
			\EndIf
		\EndFor
	\EndFor
	\State \textbf{return} true
\end{algorithmic}
\end{algorithm}

\vspace{-1.25em}
\begin{algorithm}[h]
\small
\caption{\textbf{(\textsection \ref{sec:lash-pat-algorithm}) ConfigureFallbackBaseRouting}}
\label{alg:configure-fallback-base-routing}
\begin{algorithmic}[1]
\Require{n\_routers}
\For{src = 1 to n\_routers in random order}
    \For{dst = 1 to n\_routers in random order}
        \State \parbox[t]{0.85\linewidth}{
            Get a path from src to dst using a turn-based fallback routing function. If the function provides path diversity, select a single deterministic path. Configure this path in all routing tables and add the used turns to the CDG.
        }
    \EndFor
\EndFor
\end{algorithmic}
\end{algorithm}

\vspace{-1.25em}
\begin{algorithm}[h!]
\small
\caption{\textbf{(\textsection \ref{sec:lash-pat-algorithm}) ImproveFallbackBaseRouting}}
\label{alg:improve-fallback-base-routing}
\begin{algorithmic}[1]
\Require{n\_routers}
\State paths\_improved = true
\While{paths\_improved}
	\State paths\_improved = false
	\For{src = 1 to n\_routers in random order}
		\For{dst = 1 to n\_routers in random order}
			\State len = length of the current path from src to dst
			\State success = FindPath(src, dst, len - 1)	\Comment{Alg. \ref{alg:find-path}}
			\If{success}
				\State paths\_improved = true
				\State \parbox[t]{0.75\linewidth}{Remove the old path from src to dst of length len from the routing tables and from the CDG}
			\EndIf
		\EndFor
	\EndFor
\EndWhile
\end{algorithmic}
\end{algorithm}

\newpage

\begin{algorithm}[h]
\small
\caption{\textbf{(\textsection \ref{sec:lash-pat-algorithm}) InsertAdditionalPaths}}
\label{alg:insert-additional-paths}
\begin{algorithmic}[1]
\Require{n\_routers}
	\State paths\_inserted = false
	\For{cur = 1 to n\_routers in random order}
		\For{dst = 1 to n\_routers in random order}

			\Comment{\parbox[t]{0.86\linewidth}{the loop below iterates over all (in\_port, in\_vc) pairs for which the routing table of the cur router has an entry}}
			\For{in\_chan in routing\_tables[cur][dst]}
				\State len = length of path routing\_tables[cur][dst][in\_chan]
				\State success = FindPath(cur, dst, len, in\_chan)			\Comment{Alg. \ref{alg:find-path}}
				\State paths\_inserted = paths\_inserted $\lor$ success
			\EndFor
		\EndFor
	\EndFor
	\State \textbf{return} paths\_inserted
\end{algorithmic}
\end{algorithm}

\vspace{-1.5em}
\begin{algorithm}[h]
\small
\caption{\textbf{(\textsection \ref{sec:lash-pat-algorithm}) FindPath}}
\label{alg:find-path}
\begin{algorithmic}[1]
\Require{cur, dst, max\_len, in\_chan = None}
	\State \parbox[t]{1.0\linewidth}{
    Use a modified version of Dijkstra's algorithm to find the cheapest path from cur to dst. The four main modifications from standard Dijkstra's algorithm are:
    \\
    (i) we only allow paths that do not induce any cycles in the CDG.
    The argument in\_chan is used to identify the incoming channel at the cur router.
	This channel depends on the first channel of the path, and this dependency is considered when checking for cycles in the CDG.
    If in\_chan is None, then we are looking for a path for packets that are injected at the cur router, and thus there are no dependencies to consider for the first channel of the path. 
    \\
    (ii) we use a cost function to assign a cost to each channel. The cost is in the range $[1, 1 + (1/\text{n\_routers})]$, which ensures that $\lfloor cost(path) \rfloor$ is equal to the length of the path.
    \\ 
    (iii) the first channel of the path must be different from all channels already in routing\_tables[cur][dst][in\_chan] since we want to provide path diversity, not store the same path twice.
    \\ 
    (iv) the search does not necessarily have to reach the dst router; instead, it is sufficient to find a path that reaches any other router that already has at least one path to dst in its routing table.
	}\vspace{0.2em}
    \State len = $\lfloor cost(path) \rfloor$ for the path found by Dijkstra's algorithm.
    \If{len $\leq$ max\_len} 
        \State Configure the new path in the routing tables and add the corresponding edges to the CDG.
        \State \textbf{return} true
    \Else
        \State \textbf{return} false
    \EndIf
\end{algorithmic}
\end{algorithm}
\vspace{-0.75em}

\underline{In the second step}, we insert additional paths (see \Cref{alg:insert-additional-paths}), implementing our key mechanism \gray{\textbf{M1}}.
This step is not performed per source-destination pair, but rather for each existing entry in the routing table.
This is important because providing multiple paths only at the source would not be sufficient to evade congestion that occurs along the path.
For each routing table entry currently present, we try to find additional paths that leave the current router over a different outgoing port and \gls{vc}, but eventually reach the same destination or converge with an existing path that leads to the same destination.
We only allow paths that are as short as the existing path found during the first step and that do not create cycles in the \gls{cdg} over \glspl{vc} (\gray{\textbf{M2}}).
This procedure is repeated for a configurable number of iterations, and a detailed analysis of the impact of this parameter is provided in \Cref{sec:lash-pat-n-additional-paths}.
Note that \name might fail to find the configured number of additional paths for some routing table entries, hence, this parameter is a maximum number of additional paths, not a guaranteed number of additional paths.
During this second step, we no longer use a cost function that strictly minimizes the number of edges in the \gls{cdg}.
Instead, we evaluate different cost functions (see \Cref{sec:lash-pat-cost-function}) and conclude that minimizing the number of edges in the \gls{cdg} while concurrently balancing paths across physical and virtual channels yields the highest throughput.

\subsection{An Example to Illustrate the LASH-PAT Algorithm}
\label{sec:lash-pat-example}

\begin{figure*}[h]
\centering
\captionsetup{justification=centering}
\includegraphics[width=1.0\textwidth]{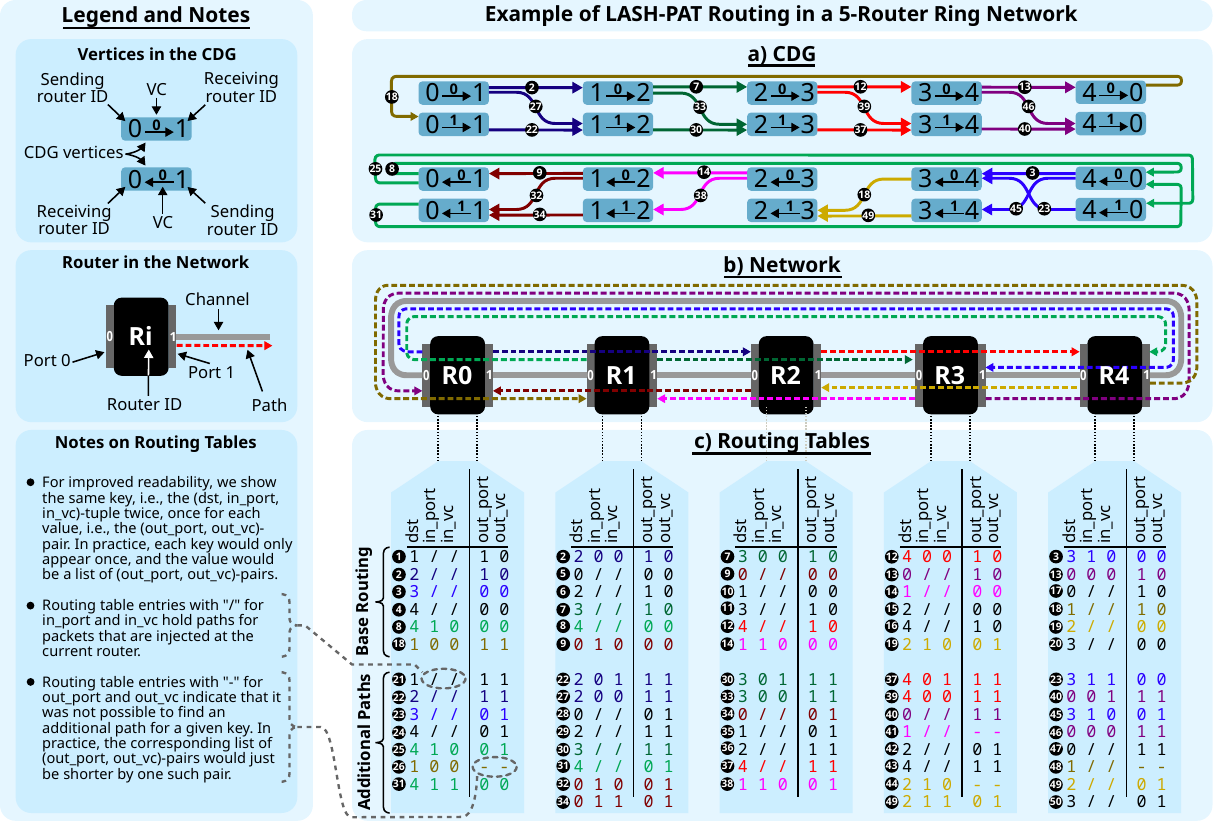}
\vspace{-1.5em}
	\caption{{(\textsection \ref{sec:lash-pat-example}) \textbf{An example of LASH-PAT routing} in a 5-router ring network.}}
\vspace{-1.75em}
\label{fig:lash-pat-example}
\end{figure*}

To illustrate the LASH-PAT algorithm, we walk through the example of computing routing tables for a 5-router ring network with 2 \glspl{vc} per physical channel.
These details are intended for readers interested in a complete understanding of the algorithm and its implementation, and are not necessary for understanding the key ideas or the evaluation results.
\Cref{fig:lash-pat-example} shows the network in part b), the corresponding \gls{cdg} in part a), and the routing tables for each router in part c).
Note that the \gls{cdg} has four vertices per physical channel, one per direction and \gls{vc}.
This small ring topology has a unique shortest path for each source-destination pair, which means that the different paths differ only in their \gls{vc} assignment.
This is rare in practice but simplifies this example.
We use one distinct color for each source-destination path of length larger than 1, and black for paths of length 1, which do not induce any edges in the \gls{cdg}.
The step of the LASH-PAT algorithm in which a given \gls{cdg} edge or routing table entry is created is indicated with a number in a black circle.
We now describe step-by-step how the custom base routing and one iteration of inserting additional paths are performed.

Steps \rone~to \rtwenty~show the custom base routing.
We start with router R0 and configure paths to routers R1 to R4 in that order.
In steps \rone~and \rfour, we configure one-hop paths to direct neighbors R1 and R4, which do not induce any edges in the \gls{cdg}.
In steps \rtwo~and \rthree, we configure two-hop paths to R2 and R3, which induce edges in the \gls{cdg} and require configuring the second hop in routers R1 and R4, respectively.
For these paths, we can use any \gls{vc} since none of the shortest paths creates cycles in the \gls{cdg}.
For routers R1 to R3, we proceed in the same way, corresponding to steps \rfive~to \rsixteen.
When configuring the paths from R4, we encounter the first situation where the \gls{cdg} could become cyclic.
While the one-hop paths to R0 and R3 configured in steps \rseventeen~and \rtwenty~do not induce any edges in the \gls{cdg}, the two-hop paths to R1 configured in step \reighteen~and to R2 configured in step \rnineteen~require special attention.
In step \reighteen, we can use any \gls{vc} to route from R4 to R0; however, we cannot proceed from R0 to R1 on \gls{vc} 0 because this would create a cycle in the \gls{cdg}.
Instead, we must use \gls{vc} 1 for the second hop of this path.
The same situation arises when routing from R4 to R2 via R3 in step \rnineteen.

In steps \rtwentyone~to \rfifty, we configure one additional path for each existing routing table entry.
For clarity, we draw them as separate entries in \Cref{fig:lash-pat-example}c), but in practice each key appears only once in a routing table, and its value is a list of next hops.
For router R0, in steps \rtwentyone~to \rtwentyfour, we configure alternative paths using \gls{vc} 1 to those configured in steps \rone~to \rfour, which use \gls{vc} 0.
In step \rtwentytwo, the path to R2 continues on \gls{vc} 1 after R1 since converging with the existing path on \gls{vc} 0 would create a cycle in the \gls{cdg}.
In step \rtwentythree, the path to R3 converges with the existing path on \gls{vc} 0 after R4 since this does not create a cycle in the \gls{cdg}.
In steps \rtwentyfive~and \rtwentysix, we attempt to configure alternatives for the paths configured in steps \reight~and \reighteen.
These are the steps in which we add diverging paths, which allow the selection function to avoid congestion along the path, as opposed to the alternative paths at the source configured in steps \rtwentyone~to \rtwentyfour.
In step \rtwentyfive, we seek an alternative path from R0 to R4 for packets arriving at port 1 and \gls{vc} 0.
In addition to the existing hop via port 0 and \gls{vc} 0, we can also route via port 0 and \gls{vc} 1, hence the alternative path can be configured.
In step \rtwentysix, we seek an alternative path from R0 to R1 for packets arriving at port 0 and \gls{vc} 0.
Since the only shortest path leads directly to R1, and \gls{vc} 1 of that hop is already used while \gls{vc} 0 is not usable as it would induce a cycle in the \gls{cdg}, we cannot configure an alternative path for this routing table entry, indicated by the entry \texttt{- -} in the routing table of R0 in step \rtwentysix.
Configuring the remaining alternative paths in steps \rtwentyseven~to \rfifty~proceeds in a similar way.

\subsection{Selecting the Number of Additional Paths}
\label{sec:lash-pat-n-additional-paths}

\begin{figure}[h]
\captionsetup{justification=centering}
\center
\includegraphics[width=0.95\columnwidth]{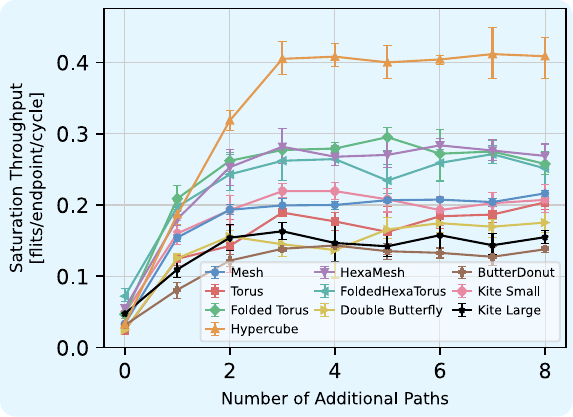}
	\caption{{(\textsection \ref{sec:lash-pat-n-additional-paths}) \textbf{Influence of the number of additional paths on throughput} for random permutation traffic.}}
\label{fig:n-additional-paths}
\vspace{-2.0em}
\end{figure}

\begin{figure*}[h]
\centering
\captionsetup{justification=centering}
\includegraphics[width=1.0\textwidth]{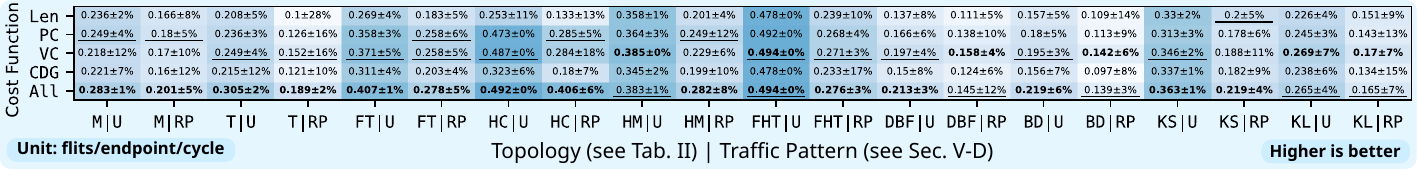}
	\caption{{(\textsection \ref{sec:lash-pat-cost-function}) \textbf{Saturation throughput of LASH-PAT for different cost functions, topologies, and traffic patterns}.}}
\vspace{-1.5em}
\label{fig:cost-function}
\end{figure*}


\Cref{fig:n-additional-paths} shows the saturation throughput of LASH-PAT in ten different network topologies when using only the base routing (zero additional paths, i.e., without \gray{\textbf{M1}}) and when using one to eight additional paths.
The experimental setup is identical to that described in \Cref{sec:eval-setup}, and results are averaged over 5 runs with error bars showing $95\%$ \glspl{ci}.
We observe that adding one additional path more than doubles the throughput, and a second additional path yields a further increase.
Depending on the network topology, the point of diminishing returns is reached between two and four additional paths, after which further paths no longer improve throughput.

An interesting observation is that the width of the \glspl{ci} depends on the network topology.
The variation in throughput is mainly caused by differences in the order in which LASH-PAT processes source-destination pairs.
When this order is fixed, the remaining variation due to randomness in the traffic pattern is small, yielding narrow \glspl{ci}.
Hence, in some topologies, the order in which source-destination pairs are processed has a large impact on throughput, while in other topologies it has little impact.
In practice, one could try many different orders and use the one that yields the highest throughput, but in this work, we always use five different random orders and report the average results.

The latency is largely independent of the number of additional paths, and the variation in latency due to the order in which source-destination pairs are processed is negligible.
Trends in energy per packet are inversely correlated with throughput, since higher throughput amortizes the static energy consumption of the network more effectively.
Both of these plots are omitted due to space constraints; they are available in our open-source repository: \githuburl.

\subsection{Selecting the Cost Function for our Modified Dijkstra}
\label{sec:lash-pat-cost-function}

Using the same evaluation setup as described in \Cref{sec:eval-setup}, we analyze five different cost functions for the second step of the LASH-PAT algorithm, which inserts additional paths.
\Cref{fig:cost-function} shows the saturation throughput of LASH-PAT using these cost functions across ten different network topologies (see \Cref{tab:topologies} for topology acronyms) under random uniform (U) and random permutation (RP) traffic patterns.
The cost functions are \texttt{Len}, which assigns a unit cost of 1 to each channel and therefore minimizes path length only, \texttt{PC}, which adds a small penalty proportional to the number of paths using a given physical channel in addition to the unit cost, \texttt{VC}, which adds a small penalty proportional to the number of paths using a given \gls{vc} in addition to the unit cost, \texttt{CDG}, which adds a small penalty for paths that induce new edges in the \gls{cdg}, and \texttt{All}, which combines all three penalties.

We observe that the cost function \texttt{All} yields the highest throughput in 15 out of 20 cases and the second-highest throughput in the remaining five cases; we therefore use it as the default cost function for LASH-PAT.
We also evaluate these cost functions in the first step of the LASH-PAT algorithm, which configures the base routing.
We find that using \texttt{CDG} in the first step yields the highest throughput, confirming the intuition that minimizing the number of edges in the \gls{cdg} during base routing is important for enabling additional path construction in the second step.

\section{Evaluation}
\label{sec:eval}

We evaluate the throughput, latency, and energy per packet of our proposed LASH-PAT \gls{rf} against nine baseline \glspl{rf} across ten \gls{ici} topologies, using two synthetic traffic patterns and two application workload traces.

\subsection{Baseline Routing Functions}
\label{sec:eval-baselines}

From the class of turn-based \glspl{rf}, we evaluate Prefix routing \cite{prefix-routing}, Up*/Down* \cite{updown}, Left/Right \cite{lturn}, L-Turn \cite{lturn}, and the \gls{scb} algorithm \cite{scb}.
For these \glspl{rf}, routing tables are configured to permit all shortest paths (with respect to the imposed turn restrictions), and all \glspl{vc} are available for each routing decision.
Many of these \glspl{rf} require the selection of a root router for \gls{mst} construction.
We select this root randomly, which is the main source of variation in performance across repetitions of the same experiment for these turn-based \glspl{rf}.

From the class of layered \glspl{rf}, we evaluate LASH \cite{lash}, DFSSSP \cite{dfsssp}, A-LASH \cite{a-lash}, and LASH-TOR \cite{lash-tor}.
For layered \glspl{rf}, one can either assign one layer per \gls{vc} or use fewer layers with multiple \glspl{vc} per layer.
For LASH and DFSSSP, we allow the algorithms to determine the number of layers required to achieve deadlock-free routing in a given topology, and we assign \glspl{vc} to layers in a round-robin manner, potentially mapping multiple \glspl{vc} to the same layer.
For A-LASH, we use one layer per \gls{vc} and replicate each path across as many layers as possible, following the approach described in the original A-LASH paper \cite{a-lash}.
Both LASH-TOR and our proposed LASH-PAT support fallback routing and operate with any number of layers. As shown in \Cref{sec:eval-rt-size}, LASH-TOR generally performs best with a single layer and four \glspl{vc} per layer, whereas LASH-PAT usually performs best with four layers and one \gls{vc} per layer. These configurations are therefore used throughout the evaluation.
For LASH-PAT, we configure three additional paths, use the \texttt{All} cost function, and employ \gls{scb} as the fallback \gls{rf}.

\subsection{Experiment Methodology and Evaluation Metrics}
\label{sec:eval-setup}

We use the cycle-accurate BookSim2 \cite{booksim} \gls{noc} simulator, in which we implement the baseline \glspl{rf}, our proposed LASH-PAT \gls{rf}, and an adaptive \gls{sf}. Using flit-level simulations in BookSim2, we measure zero-load latency and latency under load (in cycles) and saturation throughput (in flits per endpoint per cycle) across a range of topologies and traffic patterns.
Energy per transmitted packet is estimated as the sum of router and channel energy. Router energy is obtained from the Orion3 \cite{orion} model and scaled from 45\,nm to 7\,nm using DeepScaleTool \cite{deepscaletool}. Channel energy is computed by counting the number of flits transmitted over each channel and multiplying by the channel length and an energy-per-flit-per-mm factor.
The latency under load is measured at an injection rate of $0.1$ flits per endpoint per cycle.

Each BookSim2 experiment is repeated five times. We report arithmetic means for latency, harmonic means for throughput, and geometric means for relative performance, following established benchmarking guidelines \cite{th-bench-rules}. Nonparametric 95\% \glspl{ci} are computed via bootstrap resampling. In line plots, \glspl{ci} are shown as error bars, while in heatmaps the relative half-width (defined as half the \gls{ci} width divided by the sample mean) is reported in the corresponding cell.

In heatmaps, the best-performing \gls{rf} for each topology-traffic pair is shown in bold, and the second-best is underlined. Results are annotated with $\ddagger$ if the corresponding \gls{rf} uses non-shortest paths, and with $\dagger$ if it resorts to fallback routing, which applies only to LASH-TOR and LASH-PAT.
If a given \gls{rf} is not applicable due to insufficient VCs, we report \texttt{IVC} in the corresponding heatmap cell. 
In latency under load heatmaps, whenever a given \gls{rf} does not reach the target injection rate of $0.1$ flits per endpoint per cycle, we report \texttt{ITP} (insufficient throughput) in the corresponding cell.

\subsection{Simulation Parameters}
\label{sec:eval-params}

We use $64$-bit flits, four-flit packets, four \glspl{vc} per physical channel, and an $8$-flit input buffer per \gls{vc}.
We simulate a network running at $1$\,GHz, with a router latency of four cycles and one endpoint per router (i.e., a direct network with concentration factor of one).
The channel energy is 0.1\,pJ/bit/mm \cite{energy-ref}, and the channel latency is one cycle per 2\,mm, modeling a register buffer every 2\,mm, with two-cycle \gls{phy} latency for the \gls{phy} in the sending and receiving chiplet.

\subsection{Evaluated Workloads}
\label{sec:eval-workloads}

We evaluate two synthetic traffic patterns: uniform (\texttt{U}), which models all-to-all communication as observed in \gls{moe} training \cite{uniform-ref}, and random permutation (\texttt{RP}), which reflects shuffle-style communication patterns such as those in FFT or sorting algorithms \cite{randperm-ref}.
In addition to synthetic traffic, we assess CPU and GPU workloads using traces truncated to 100K events, as simulating millions of events in BookSim2 would be prohibitively expensive in terms of runtime.
For CPU workloads, we use Netrace \cite{netrace,netrace-tr} with the \textit{blackscholes} trace \cite{netrace-traces}, representing a PDE-based financial application that computes option prices using the Black-Scholes equation \cite{black-scholes}.
For GPU workloads, we rely on the ATLAHS \cite{atlahs} toolchain to generate GOAL \cite{goal} traces from training the \textit{Llama-7B} \cite{llama} large language model \cite{llm-survey}.

\subsection{Evaluated Network Topologies}
\label{sec:eval-topologies}

We evaluate on established and emerging \gls{ici} topologies \cite{folded-torus,hypercube,hexamesh,butterdonut,double-butterfly,kite,foldedhexatorus} with either $64$ or $61$ routers (HexaMesh and FoldedHexaTorus only support $61$ routers).
Details on the topologies are provided in \Cref{tab:topologies}.

\begin{table}[h]
\small
\setlength{\tabcolsep}{7.5pt}
\centering
\captionsetup{justification=centering}
\caption{(\textsection \ref{sec:eval-topologies}) \textbf{Overview of investigated ICI topologies}.}
\vspace{-0.5em}
\begin{tabular}{lccc}
\toprule
\textbf{\makecell[l]{Topology}}
& \rotatebox[origin=l]{0}{\textbf{\makecell[c]{\hspace{-0.6em}Number of\\\hspace{-0.6em}Routers}}}
& \rotatebox[origin=l]{0}{\textbf{\makecell[c]{\hspace{-0.6em}Network\\\hspace{-0.6em}Diameter}}}
& \rotatebox[origin=l]{0}{\textbf{\makecell[c]{\hspace{-0.6em}Router\\\hspace{-0.6em}Radix}}}
\\
\midrule
	Mesh (M)														& 64         & 14         & 2-4        \\
	Torus (T)														& 64         & 8          & 4          \\
	Folded Torus (FT) \cite{folded-torus}							& 64         & 8          & 4          \\
	HexaMesh (HM) \cite{hexamesh}                  					& 61         & 8          & 3-6        \\
	FoldedHexaTorus (FHT) \cite{foldedhexatorus}  \hspace{-2em}		& 61         & 5          & 6          \\
	Hypercube (HC) \cite{hypercube}                 				& 64         & 6          & 12         \\
	Double Butterfly (DBF) \cite{double-butterfly} \hspace{-2em}	& 64         & 8          & 2-4        \\
	ButterDonut (BD) \cite{butterdonut}             				& 64         & 6          & 3-4        \\
	Kite Small (KS) \cite{kite}                						& 64         & 7          & 3-4        \\
	Kite Large (KL) \cite{kite}                						& 64         & 8          & 2-4        \\
\bottomrule
\end{tabular}
\label{tab:topologies}
\vspace{-1.5em}
\end{table}

\subsection{Results and Discussion}
\label{sec:eval-results}

\begin{figure*}[h]
\centering
\captionsetup{justification=centering}
\includegraphics[width=1.0\textwidth]{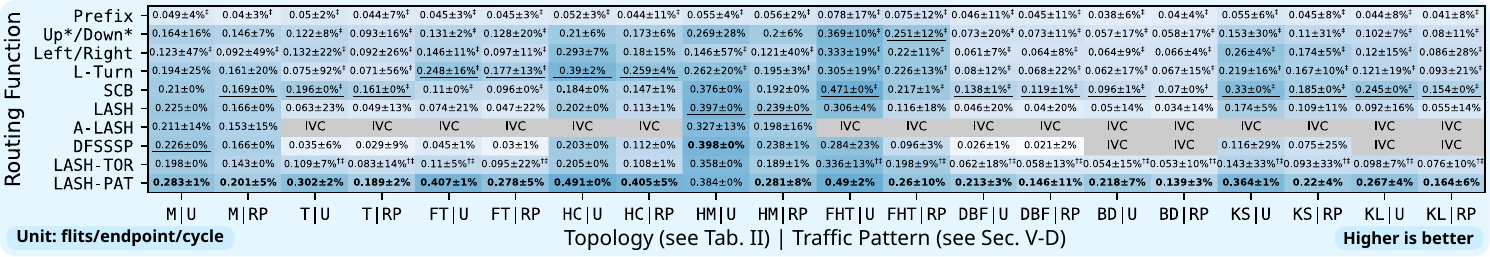}
\vspace{-1.5em}
	\caption{{(\textsection \ref{sec:eval-results-tp}) \textbf{Saturation throughput (synthetic traffic)}. $^\dagger$Fallback routing was used. $^\ddagger$Some non-shortest paths were used.}}
\label{fig:eval-tp}
\end{figure*}

\begin{figure*}[h]
\centering
\captionsetup{justification=centering}
\includegraphics[width=1.0\textwidth]{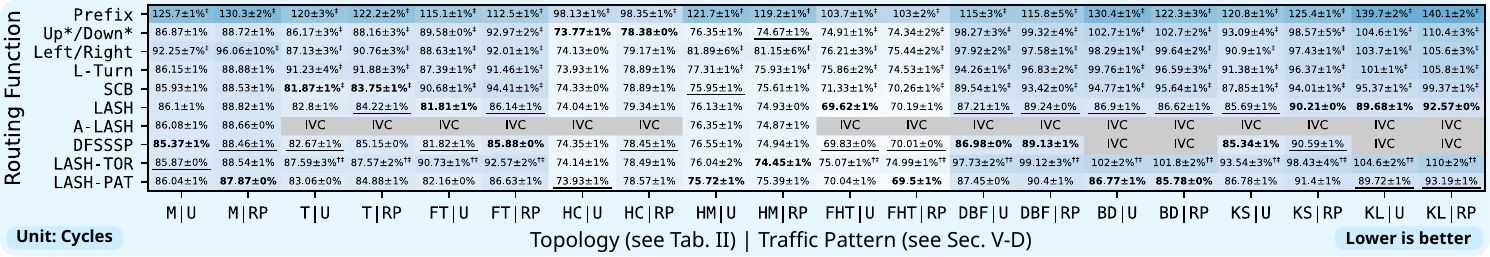}
\vspace{-1.5em}
\caption{{(\textsection \ref{sec:eval-results-lat-zero}) \textbf{Zero-load latency (synthetic traffic)}. $^\dagger$Fallback routing was used. $^\ddagger$Some non-shortest paths were used.}}
\label{fig:eval-lat-zero}
\end{figure*}

\begin{figure*}[h]
\centering
\captionsetup{justification=centering}
\includegraphics[width=1.0\textwidth]{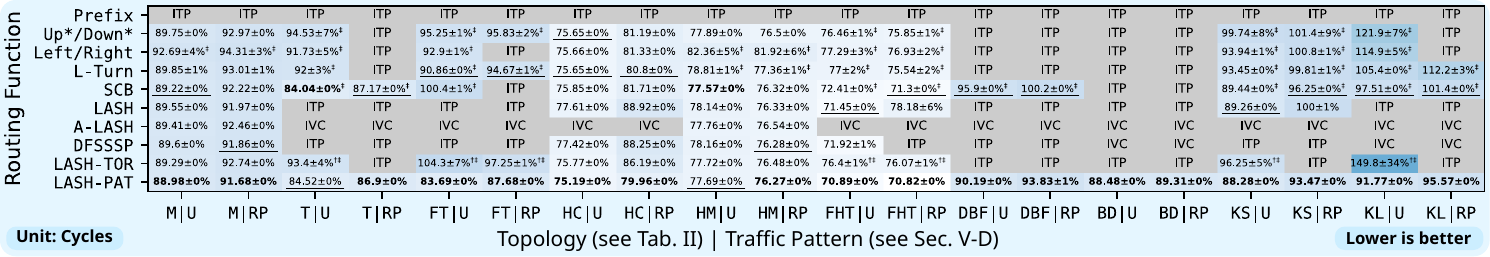}
\vspace{-1.5em}
\caption{{(\textsection \ref{sec:eval-results-lat-load}) \textbf{Latency under load (synthetic traffic)}. $^\dagger$Fallback routing was used. $^\ddagger$Some non-shortest paths were used.
	\texttt{ITP} denotes that a given RF did not sustain the 0.1 flits/endpoint/cycle throughput that was used to determine the latency under load.}}
\label{fig:eval-lat-load}
\end{figure*}

\begin{figure*}[h]
\centering
\captionsetup{justification=centering}
\includegraphics[width=1.0\textwidth]{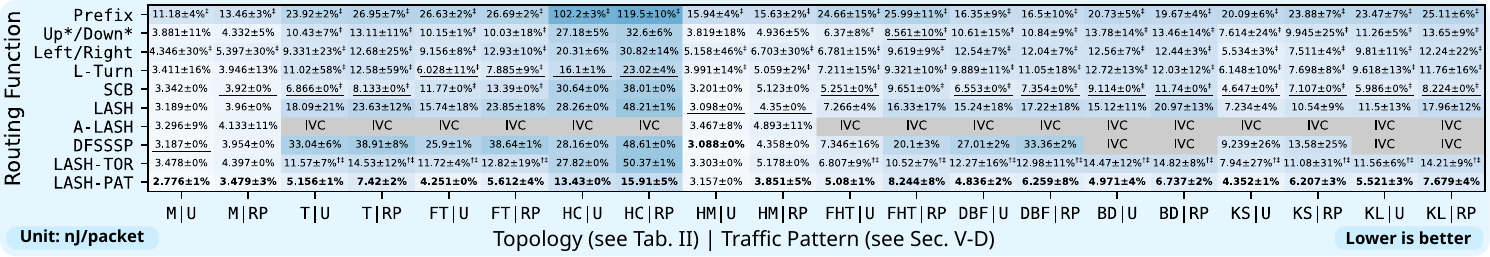}
\vspace{-1.5em}
\caption{{(\textsection \ref{sec:eval-results-energy}) \textbf{Energy per packet (synthetic traffic)}. $^\dagger$Fallback routing was used. $^\ddagger$Some non-shortest paths were used.}}
\label{fig:eval-energy}
\end{figure*}

\begin{figure*}[h!]
\centering
\captionsetup{justification=centering}
\includegraphics[width=1.0\textwidth]{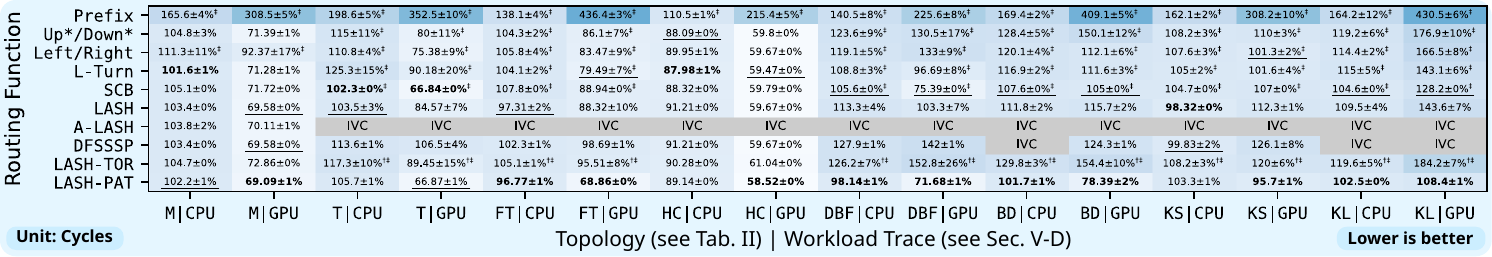}
\vspace{-1.5em}
	\caption{{(\textsection \ref{sec:eval-results-trace}) \textbf{Average packet latency (traffic trace)}. $^\dagger$Fallback routing was used. $^\ddagger$Some non-shortest paths were used. \\
	Note that HexaMesh and FoldedHexaTorus were excluded from this experiment because they use $61$ routers, whereas our traces use $64$ endpoints.}}
\label{fig:eval-trace}
\vspace{3em}
\end{figure*}

\subsubsection{Throughput}
\label{sec:eval-results-tp}

\Cref{fig:eval-tp} shows the saturation throughput of all evaluated \glspl{rf} across all topologies and the two synthetic traffic patterns.
LASH-PAT achieves the highest throughput in 19 out of 20 topology-traffic pairs.
\gls{scb} or L-Turn, both of which are turn-based \glspl{rf} providing path diversity, frequently take second place.
This observation confirms that path diversity is crucial for achieving high throughput, which is the primary motivation for extending layered routing with path diversity (key mechanism \gray{\textbf{M1}}) in LASH-PAT.
\Cref{tab:improvement} reports the percentage throughput improvement of LASH-PAT over each baseline \gls{rf}.
Averaged across all topology-traffic pairs, LASH-PAT improves throughput by 31.92\% over \gls{scb}, the best turn-based \gls{rf}, and by 105.61\% over LASH-TOR, the best layered \gls{rf} applicable to all topologies at $4$ \glspl{vc}.
When compared to the baseline \glspl{rf} achieving the highest throughput in each specific topology-traffic pair, LASH-PAT still provides a 21.15\% average throughput increase.

\begin{table}[h]
\vspace{-0.5em}
\footnotesize
\setlength{\tabcolsep}{5.5pt}
\centering
\captionsetup{justification=centering}
\caption{(\textsection \ref{sec:eval-results}) \textbf{Performance improvement of LASH- PAT over baselines}. Each value is averaged over all topolo- gies and traffic patterns. The row labeled ''Best`` compares against the best-performing RF for each topology-traffic pair.}
\vspace{-0.5em}
\begin{tabular}{llllll}
\toprule
	\textbf{\makecell[l]{Baseline\\Routing\\Function\\Compared\\Against}}
	& \rotatebox[origin=l]{90}{\textbf{\makecell[l]{\hspace{-2.6em}Saturation\\\hspace{-2.6em}Throughput\\\hspace{-2.6em}(Synth. Traffic)}}}
	& \rotatebox[origin=l]{90}{\textbf{\makecell[l]{\hspace{-2.6em}Zero-Load\\\hspace{-2.6em}Latency\\\hspace{-2.6em}(Synth. Traffic)}}}
	& \rotatebox[origin=l]{90}{\textbf{\makecell[l]{\hspace{-2.6em}Latency\\\hspace{-2.6em}Under Load\\\hspace{-2.6em}(Synth. Traffic)}}}
	& \rotatebox[origin=l]{90}{\textbf{\makecell[l]{\hspace{-2.6em}Energy\\\hspace{-2.6em}per Packet\\\hspace{-2.6em}(Synth. Traffic)}}}
	& \rotatebox[origin=l]{90}{\textbf{\makecell[l]{\hspace{-2.6em}AVG Packet\\\hspace{-2.6em}Latency\\\hspace{-2.6em}(Traffic Trace)}}}
\\
\midrule
Prefix      & \gn{+444.55\%}    & \gn{-29.33\%}     & N/A      			& \gn{-75.36\%}    & \gn{-50.63\%}      \\
Up*/Down*   & \gn{+86.78\%}     & \gn{-3.77\%}      & \gn{-2.44\%}      & \gn{-34.63\%}    & \gn{-10.08\%}      \\
Left/Right  & \gn{+100.17\%}    & \gn{-5.94\%}      & \gn{-4.39\%}      & \gn{-37.75\%}    & \gn{-9.88\%}       \\
L-Turn      & \gn{+68.72\%}     & \gn{-2.7\%}       & \gn{-1.78\%}      & \gn{-29.0\%}    & \gn{-6.98\%}       \\
SCB         & \gn{+31.92\%}     & \gn{-1.71\%}      & \gn{-1.28\%}      & \gn{-16.61\%}    & \gn{-3.7\%}        \\
LASH        & \gn{+124.59\%}    & \rd{+0.48\%}      & \gn{-0.59\%}      & \gn{-38.36\%}    & \gn{-5.98\%}       \\
A-LASH      & \gn{+29.58\%}     & \gn{-0.39\%}      & \gn{-0.39\%}      & \gn{-14.77\%}    & \gn{-1.5\%}        \\
DFSSSP      & \gn{+151.59\%}    & \rd{+0.61\%}      & \gn{-0.51\%}      & \gn{-38.58\%}    & \gn{-8.0\%}        \\
LASH-TOR    & \gn{+105.61\%}    & \gn{-3.33\%}      & \gn{-1.84\%}      & \gn{-38.4\%}    & \gn{-12.2\%}      \\
\midrule
\textbf{Best}&\gn{+21.15\%}   	& \rd{+0.59\%}      & \gn{-1.33\%}      & \gn{-12.77\%}    & \gn{-2.56\%}      \\
\bottomrule
\end{tabular}
\vspace{-1.0em}
\label{tab:improvement}
\end{table}

\begin{figure*}[h]
\centering
\captionsetup{justification=centering}
\includegraphics[width=1.0\textwidth]{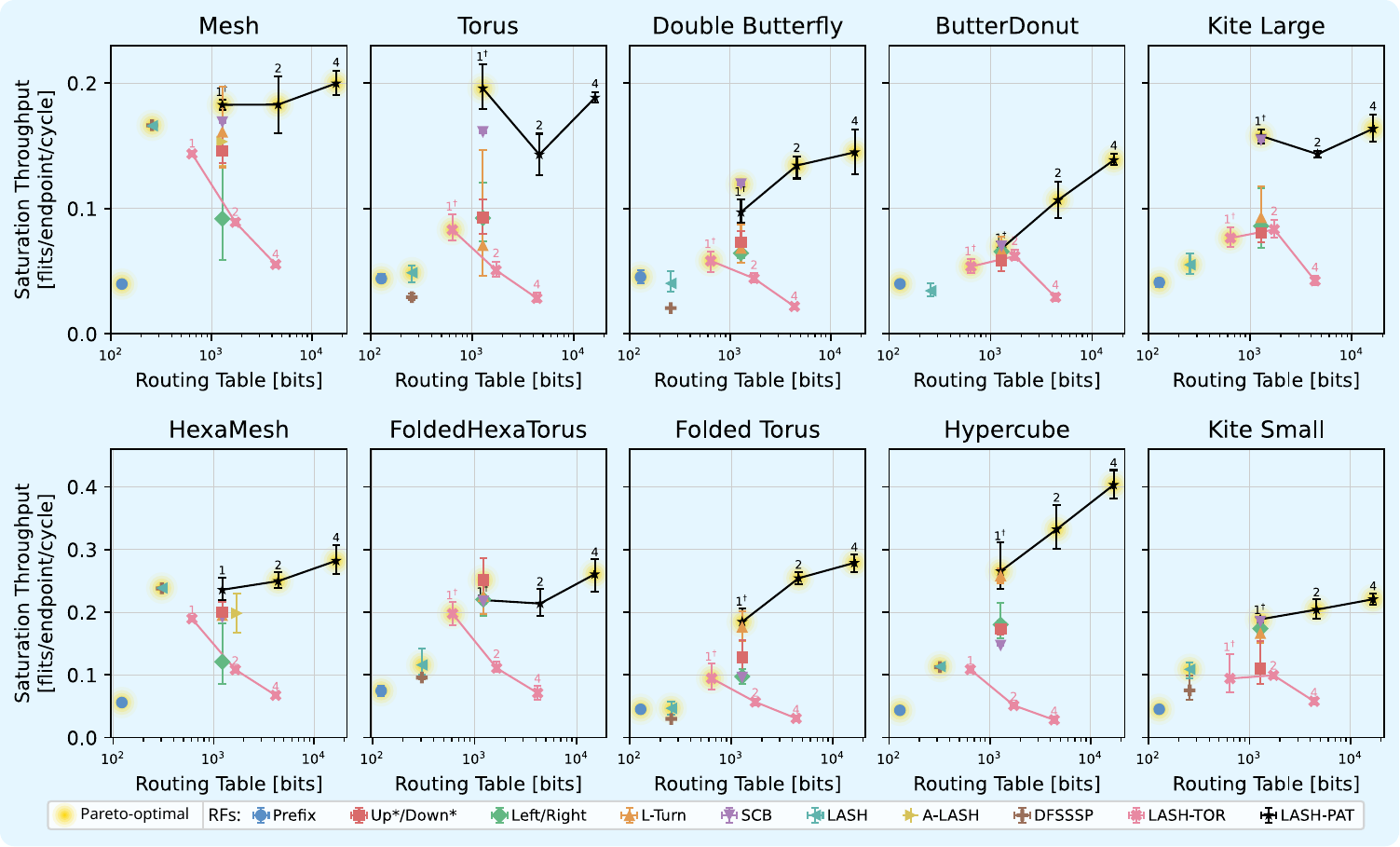}
\vspace{-1.5em}
\caption{{(\textsection \ref{sec:eval-rt-size}) \textbf{Throughput vs. routing table size}. Numbers indicate virtual layer counts. $^\dagger$Fallback routing was used.}}
\label{fig:eval-rt-size}
\vspace{-1.5em}
\end{figure*}


\subsubsection{Zero-Load Latency}
\label{sec:eval-results-lat-zero}

\Cref{fig:eval-lat-zero} reports the zero-load latency.
As long as shortest paths are used, zero-load latency depends primarily on the topology and remains largely independent of the \gls{rf}.
\glspl{rf} generating a large number of non-shortest paths, such as Prefix routing, incur a significant increase in zero-load latency.
This confirms that shortest-path routing is essential for low-latency communication, motivating the combination of adaptive and shortest-path routing in LASH-PAT.
LASH-PAT achieves low zero-load latency, performing only marginally worse than LASH (+0.48\%) and DFSSSP (+0.61\%).

\subsubsection{Latency Under Load}
\label{sec:eval-results-lat-load}

The trends in latency under load (see \Cref{fig:eval-lat-load}) mirror those in zero-load latency, except that at an injection rate of $0.1$ flits per endpoint per cycle, LASH-PAT achieves the lowest latency among all evaluated \glspl{rf}.

\subsubsection{Energy per Packet}
\label{sec:eval-results-energy}

\Cref{fig:eval-energy} illustrates the energy per packet.
LASH-PAT achieves the lowest energy per packet in 19 out of 20 topology-traffic pairs, reducing energy per packet by 12.77\% on average compared to the most energy-efficient baseline \gls{rf} in each respective pair.

\subsubsection{Application Workload Traces}
\label{sec:eval-results-trace}

\Cref{fig:eval-trace} shows the average packet latency under two application workload traces, one between CPU chiplets and the other between GPU chiplets.
In contrast to synthetic traffic, where latency was very similar as long as shortest paths were used, application workloads exhibit larger latency differences between \glspl{rf}.
LASH-PAT achieves the lowest latency in 11 out of 16 topology-trace pairs, beating the second-best \gls{rf} by up to 25.3\% (on the ButterDonut topology with the GPU trace).

\subsubsection{Observations on LASH-PAT}
\label{sec:eval-results-observation}

As described in \Cref{sec:lash-pat-algorithm}, LASH-PAT's custom base routing may configure non-shortest paths if the available \glspl{vc} are insufficient to support all shortest paths while guaranteeing deadlock freedom, or it may resort to fallback routing if the custom base routing fails.
Our results in \Cref{fig:eval-tp,fig:eval-lat-zero,fig:eval-lat-load,fig:eval-energy,fig:eval-trace} show that across all ten evaluated topologies, LASH-PAT consistently establishes shortest paths without resorting to non-shortest paths or fallback routing.
Together with the fact that A-LASH fails in eight out of ten topologies due to insufficient \glspl{vc} and DFSSSP fails in two out of ten, this highlights the benefit of LASH-PAT's fine-grained assignment of path segments to virtual layers via key mechanism \gray{\textbf{M2}}, enabling deadlock-free shortest-path routing even in complex topologies with a limited number of \glspl{vc}.

\subsection{Discussion on Routing Table Size}
\label{sec:eval-rt-size}

\begin{figure*}[h]
\centering
\captionsetup{justification=centering}
\includegraphics[width=1.0\textwidth]{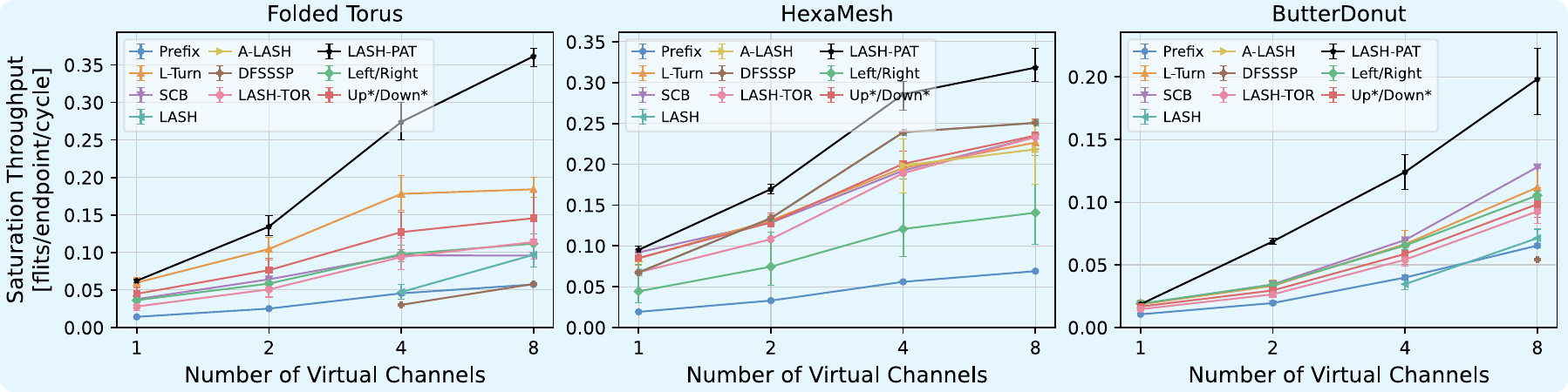}
\vspace{-1.5em}
	\caption{{(\textsection \ref{sec:eval-vcs}) \textbf{Throughput of ten RFs for varying VCs}. Missing data points indicate that an RF has insufficient VCs.}}
\label{fig:vary-vcs-tp}
\vspace{-1.0em}
\end{figure*}

\begin{figure*}[h]
\centering
\captionsetup{justification=centering}
\includegraphics[width=1.0\textwidth]{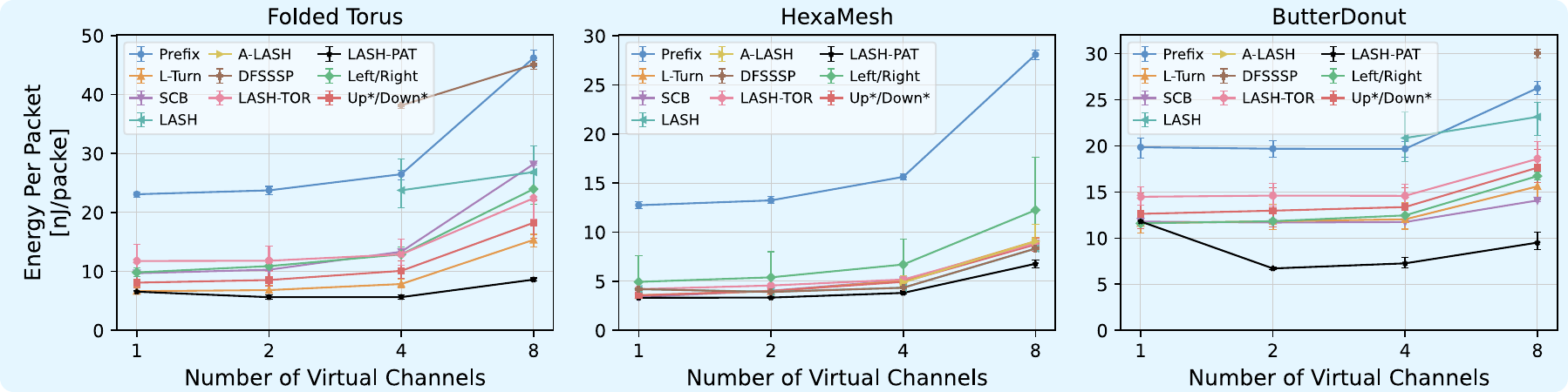}
\vspace{-1.5em}
\caption{{(\textsection \ref{sec:eval-vcs}) \textbf{Energy per packet of 10 RFs for varying VCs}. Missing data points indicate that a RF has insufficient VCs.}}
\label{fig:vary-vcs-energy}
\vspace{-1.5em}
\end{figure*}


Routing table size depends on both the number of entries and the size of each entry in the routing table.
\glspl{rf} indexing routing tables solely by the destination router (Prefix routing, LASH, A-LASH, DFSSSP) require fewer entries than those also depending on the input port (Up*/Down*, Left/Right, L-Turn, \gls{scb}).
In turn, these require fewer entries than \glspl{rf} additionally indexed by the input \gls{vc} (LASH-TOR, LASH-PAT).
Entry size depends on whether the table stores a single output (Prefix routing, LASH, LASH-TOR, DFSSSP) or a list of outputs (A-LASH, Up*/Down*, Left/Right, L-Turn, \gls{scb}, LASH-PAT).
For layered \glspl{rf}, using the layer ID instead of the \gls{vc} ID in keys and values reduces the routing table size when fewer layers than \glspl{vc} are utilized.

\Cref{fig:eval-rt-size} plots saturation throughput against routing table size for all evaluated \glspl{rf} across all topologies under random permutation traffic.
For LASH-TOR and LASH-PAT, we vary the number of layers from one (four \glspl{vc} per layer) to four (one \gls{vc} per layer) and annotate the corresponding data points with the number of layers.
We observe that LASH-PAT's superior throughput is achieved at the expense of a larger routing table, a trade-off adjustable by varying the layer count.
Interestingly, while LASH-PAT generally improves with more layers, LASH-TOR performs best with a single layer and deteriorates as the layer count increases.
Because LASH-TOR provides a single deterministic path per source-destination pair, using fewer layers with multiple \glspl{vc} enables traffic balancing across \glspl{vc} and effectively multiplies the available buffer space along the path.
Consequently, we evaluate LASH-TOR using one layer, which delivers the best performance despite forcing frequent reliance on fallback routing and non-shortest paths.
Note that for all topologies, LASH-PAT yields at least one, and in many cases even three, Pareto-optimal \cite{pareto-optimality} configurations (highlighted in yellow in \Cref{fig:eval-rt-size}).

Large routing tables are typically problematic for \glspl{noc} in monolithic chips due to strict area budgets and for datacenter networks due to the necessity of scaling to thousands of endpoints.
However, we argue that neither of these limitations applies to \glspl{ici}.
Unlike datacenter environments, the total number of interconnected chiplets remains relatively small, with our evaluated 64-chiplet topologies being on the higher end of the spectrum.
Furthermore, unlike highly area-constrained \gls{noc} routers, individual chiplets provide massive silicon real estate, reaching $74~\text{mm}^2$ in AMD’s EPYC and Ryzen processors \cite{amd-chiplets} and approaching the reticle limit with $\sim$645\,mm$^2$ in the wafer-scale DOJO chip \cite{dojo}.
\Cref{fig:eval-rt-size} demonstrates that the worst-case routing table size for LASH-PAT approaches 20~Kbit.
This capacity occupies roughly $0.003~\text{mm}^2$ in a 22\,nm process and $0.001~\text{mm}^2$ in a 7\,nm process, yielding an area footprint that is entirely negligible relative to the total chiplet area.

\subsection{Ablation Study on the Number of VCs}
\label{sec:eval-vcs}

\Cref{fig:vary-vcs-tp,fig:vary-vcs-energy} evaluate throughput and energy per packet across varying \gls{vc} counts for LASH-PAT and the nine baseline \glspl{rf}. LASH-PAT consistently achieves the highest throughput and lowest energy per packet across the three evaluated topologies, with the sole exception of ButterDonut when using a single VC.
Although throughput universally scales with additional \glspl{vc}, LASH-PAT's energy consumption initially drops as higher throughput amortizes static power and increases only at high VC counts due to larger input buffer overheads. Notably, unlike LASH, A-LASH, and DFSSSP, LASH-PAT remains compatible with all topologies regardless of VC count.

\section{Conclusion}
\label{sec:conc}

\glsresetall

In this paper, we present the first comprehensive evaluation of deadlock-free routing functions for irregular inter-chiplet interconnects.
We identify a trade-off in routing for irregular topologies: turn-based methods enable path diversity for high throughput, whereas layered methods guarantee shortest paths for low latency.
To bridge this gap, we propose \name, which extends \underline{la}yered \underline{sh}ortest-path routing with \underline{pat}h diversity (key mechanism \gray{\textbf{M1}}) and a more efficient assignment of paths to layers (key mechanism \gray{\textbf{M2}}).
When evaluated against the best-performing baseline for each scenario, \name improves throughput by 21.15\% and reduces energy per packet by 12.77\%, while providing competitive latency.
While supporting this adaptive path diversity inherently requires larger routing tables, the resulting memory overhead is practically negligible for modern chiplet-based systems.
In conclusion, \name delivers high-throughput, low-energy communication for next-generation chiplet systems.

\ifnb
\section*{Acknowledgements}
\label{sec:ack}

We thank Luca Oertig for his help with implementing several of our baseline routing functions in the course of another project.
This work was supported by the ETH Future Computing Laboratory (EFCL), financed by a donation from Huawei Technologies.
It also received funding from the European Research Council \raisebox{-0.25em}{\includegraphics[height=1em]{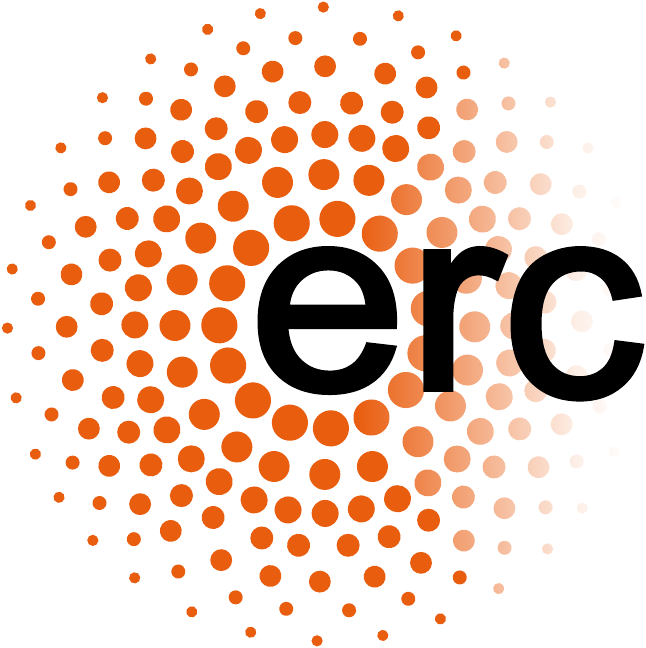}} (Project PSAP, No.~101002047).

\fi

\bibliographystyle{IEEEtranS}
\bibliography{bibliography}

\end{document}